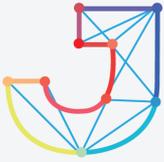
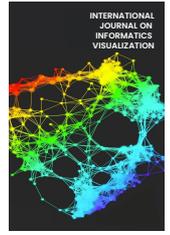

# INTERNATIONAL JOURNAL ON INFORMATICS VISUALIZATION

journal homepage : www.joiv.org/index.php/joiv

# Vulnerability Liquefaction Mapping in Padang City Based on Cloud Computing Using Optical Satellite Imagery Data

Pakhrur Razi [a,b], Amalia Putri [a,b], Josaphat Tetuko Sri Sumantyo [c], Akmam [a,b]

[a] *Physics Department, Universitas Negeri Padang, West Sumatra, Indonesia*
[b] *Center of Disaster Monitoring and Earth Observation, Physics Department, Universitas Negeri Padang, West Sumatra, Indonesia*
[c] *Center for Environmental Remote Sensing, Chiba University, Chiba, Japan*
Corresponding author: *fhrrazi@fmipa.unp.ac.id*Abstract*—Liquefaction is a significant geological hazard in earthquake-prone locations like Padang City, Indonesia. The phenomenon happens when saturated soil loses strength owing to seismic shaking, resulting in substantial infrastructure damage. Accurate identification of sensitive locations is critical to catastrophe mitigation. This study aims to map water distribution using optical satellite data and estimate its importance as a crucial element in determining liquefaction vulnerability. The Normalized Difference Water Index (NDWI) was used to assess water and vegetation indexes, taking advantage of its sensitivity to water content in varied land surfaces. We recommended using the NIR (near-infrared) and SWIR (short wave infrared) bands with 832.8 nm and 2202.4 nm, respectively, which are sensitive to soil water content. High-resolution satellite data were used to create NDWI maps, highlighting locations with high water saturation. These findings were combined with geological and seismic data to identify liquefaction-prone zones. The study found that locations with high water content, as measured by NDWI, are highly associated with greater liquefaction susceptibility. The findings highlight the importance of water distribution in determining soil behavior during seismic occurrences. This study highlights the value of NDWI as a low-cost and efficient tool for measuring liquefaction vulnerability at the regional level. The technique offers insights into Padang City's urban planning, catastrophe risk reduction, and community preparedness. By identifying high-risk zones, the study aids in making informed decisions to reduce the impact of future earthquakes. Most of the water content change occurred along the coastal line and in the low-lying areas of Koto Tanggah and North Padang sub-districts. The model can be used in other places with similar geological challenges, providing a scalable solution for liquefaction risk assessment.

*Keywords*— Liquefaction; NDVI; water content; water index; Padang.

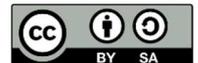

## I. INTRODUCTION

Padang City is located on low land on the coastal line of West Sumatra. In the past, most of the area along the coast of Padang City was a marsh area that grew with sago palm (Rumbia). The area has become a residential area, with office buildings and campuses for a few decades. However, the increasing number of earthquakes occurring in Padang City in the last ten years and the entire 200 years period of the super-cycle of earthquakes in the megathrust zone in the western part of Padang [1], the existence of buildings in the area needs to be evaluated [2], [3].

One of the major earthquakes in Padang was in 2009, with a magnitude of 7.6 on the Richter scale. This earthquake caused a lot of damage to the soil structure, including soil settlement, sand boil, lateral spreading, and loss of soil strength, which is known as liquefaction [4], [5]. This phenomenon is closely related to the level of soil density, soil type, and water content in Padang. To minimize the level of damage to buildings caused by earthquakes, it is necessary to observe the condition of the soil water content as one parameter of vulnerability liquefaction [6].

Normalized Differential Water Index (NDWI) is a technique to measure an area's wetness level that is correlated with soil water content [7], [8]. The method compares the near-infrared (NIR) band and short-wave infrared (SWIR) band with 832 nm and 1613 nm wavelengths, respectively [9],[10]. The wavelength in the NIR band is sensitive to the water; almost all the light in this wavelength is absorbed by water; therefore, the waterbodies appear very dark. SWIR band is sensitive to moisture; hence, it is used to monitor the water content of the soil [11].

Over the years, the NDWI technique has been spread applied in many applications, including pasture seasonal [8], water bodies [12], flood [13], and deforestation [14]. Most NDWI processing uses NIR and SWIR1 bands, but in this study, we propose to extract not only NIR and SWIR1 but also NIR and SWIR2. SWIR2 has a longer wavelength than SWIR1, so the penetration of SWIR2 is much deeper into the ground than SWIR1. It is increasing the accuracy of groundwater content measurement. Also, SWIR2 provides good contrast between different geologic rock formations and different types of vegetation as compared to SWIR1 bands [15], [16].

In 2015, the European Space Agency (ESA) launched a new optical satellite, Sentinel-2. Sentinel-2 has twelve bands with spatial resolution in 10, 20, and 60 meters. The satellite's covered area is. Furthermore, in the analysis, we provided the vegetation index value to visualize the area's condition. Thus, this research aims to map the water content in Padang City using NDWI. The results from both combinations of NIR and SWIR1 using Equation (1) and NIR and SWIR2 Equation (2) are compared. Also, water content value and vegetation index are presented in the map's grid, representing the area's condition.

### A. Study area and Satellite dataset

The study area is in Padang City, West Sumatra, Indonesia. Padang City is one of the cities orientated to the subduction zone of Indo-Australia and the Eurasia plate in the Western part of Sumatra Island. The area's elevation is 0-1853 meters above sea level, with a covered area of 694.96 km$^2$. Padang City flows by five and 16 number big and small river basins, respectively. All the rivers end on the shore of Padang City. Most of the settlements in Padang City are distributed along the coastline with elevations about 0-10 meters above sea level. The topography, bathymetry line, and river basin of Padang city are shown in Fig. 1.

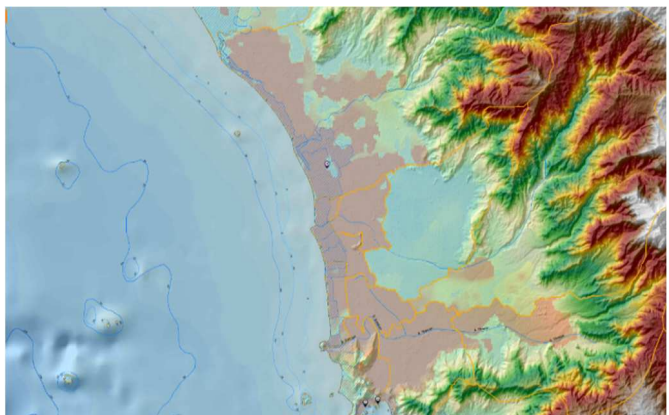

Fig. 1  Topography, bathymetry line, and river basin of Padang City

The water index in Padang city is observed using Sentinel-2 level-1C satellite product, which produced radiometric and geometric corrections, including ortho-rectification and spatial registration on a global reference system with sub-pixel accuracy [17]. The spatial resolution of Sentinel-2 product depends on the band of the satellite sensor, 10, 20, and 60 meters [18]. The Sentinel-2 dataset is listed in Table 1.

TABLE I
THE SENTINEL-2 SATELLITE DATASET

| Band Number | Spatial Resolution (m) | Central wavelength (nm) | Bandwidth (nm) |
|---|---|---|---|
| B1-Coastal aerosol | 60 | 442.7 | 21 |
| B2-Blue | 10 | 492.4 | 66 |
| B3-Green | 10 | 559.8 | 36 |
| B4-Red | 10 | 664.6 | 31 |
| B5-Vegetation Red edge | 20 | 704.1 | 15 |
| B6-Vegetation Red edge | 20 | 740.5 | 15 |
| B7-Vegetation Red edge | 20 | 782.8 | 20 |
| B8-NIR | 10 | 832.8 | 106 |
| B8A-Vegetation Red edge | 20 | 864.7 | 21 |
| B9-Water Vapor | 60 | 945.1 | 20 |
| B10-SWIR-Cirrus | 60 | 1373.5 | 31 |
| B11-SWIR | 20 | 1613.7 | 91 |
| B12-SWIR | 20 | 2202.4 | 175 |

In this study, the water content is calculated using a combination of NIR (B8) and SWIR (B11) and NIR (B8) and SWIR (B12) bands. Both types of bands have spatial resolutions of 10 and 20 meters, respectively.

## II. MATERIAL AND METHOD

### A. Normalized Difference Water Index (NDWI)

Normalized difference water Index is a technique to extract information about the water content on the earth's surface using optical satellite remote sensing. McFeeter proposed NDWI to maximize and minimize the reflection of water in the green band and near-infrared band (NIR) [9]. In Sentinel-2 data, the NDWI is processed by Modification formula using a combination of NIR band and SWIR1 [15] band, and NIR and SWIR2 band.

$$NDWI = (B08 - B11)/(B08 + B11) \quad (1)$$

$$MNDWI = (B08 - B12)/(B08 + B12) \quad (2)$$

where B08 is the Top of Atmospheric (TOA) reflectance of the Near Infrared (NIR) band and B11 and B12 are the TOA reflectance of the Short-wave infrared (SWIR) band with different wavelengths 1613.7 nm and 2202.4 nm, respectively.

The water content measurement is expressed in the water index corresponding to the soil's percentage of water. The water index value varies from -1 to +1 depending on the area's soil and vegetation type. The high-value water index (purple color) indicates the high-water content in the area. A low NDWI value (red color) corresponds to low water content. Water Index values greater than 0.5 usually correspond to water bodies. Vegetation usually corresponds to smaller values and built-up areas between zero and 0.2. The SWIR band is absorbed by water. Therefore, the index value increases with the drying of the soil.

## III. RESULTS AND DISCUSSION

The map of water content in Padang City was depicted based on the Normalized Difference Water Index (NDWI)

technique applied to Sentinel-2 data. In processing, NDWI algorithms were applied through the cloud computing platform based on Google Earth Engine (GEE) for both equations (Eq. 1 and Eq. 2). The water content value for both equations was compared and displayed on a grid map in 10 x 10 scale in each sub-district along the shoreline of Padang city. Also, we display the vegetation index value paired with the water content value in each area.

NDWI value extracted using a combination of near-infrared and short-wave infrared-2 is more sensitive to detecting the water content than the combination of near-infrared and short-wave infrared-1. So, the water content value extracted using Eq. 2 is more significant than Eq. 1. Eq. 2 SWIR causes it in B12 to have a wavelength of 2202.4 nm while the SWIR in B11 is 1613.7 nm [19]. The longer the wavelength, the deeper the penetration ability into the soil.

The significant difference in the index value of NDWI indicates the tremendous increase in the water content on the surface (red to violet), with a range of -0.355 to 0.719. A larger vegetation index suggests the area is densely covered with vegetation. The negative and positive signs of water content indicate the area is dry and wet, respectively. Furthermore, the negative and positive signs of the vegetation index indicate the area has no vegetation and dense vegetation, respectively. The water content and vegetation index map of Koto Tanggah (1) is depicted in Fig. 2.

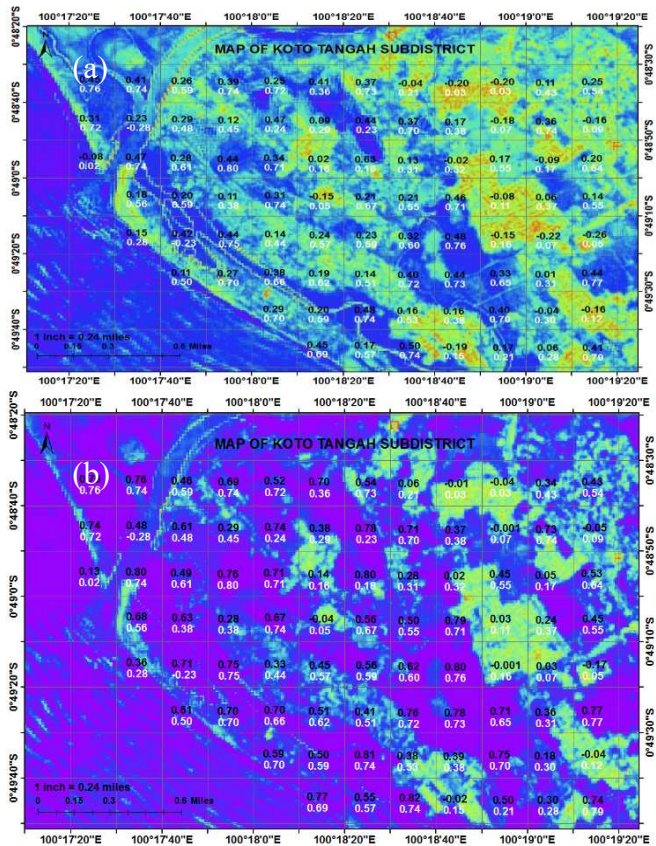

Fig. 2  The Map of water content and vegetation index in Koto Tanggah subdistrict (1): (a). NIR-SIWR1 (b). NIR-SWIR2

Fig. 2a displays the water content and vegetation index using the NDWI equation, and Fig.2b uses the MNDWI equation in the Koto Tanggah subdistrict (1). Both algorithms produced different values in each equal location. The majority are in Fig.2b with a magenta-blue color, and the residential area is yellow-red. The area with blue-magenta has a higher water content than the yellow-red area, also for the vegetation index. The area with high water content is in the coastal line, growing with grass. Most of this area is uninhabited because it is swampy and overgrown with sago palms.

Fig.3 displays the water index map of the Koto Tanggah sub-district (2). Rice fields and open spaces still dominate this area. The area along the coastline has a moderate water content with relatively sparse vegetation. However, most of the area is swamped in the eastern part, with lower elevations than the residential settlement locations.

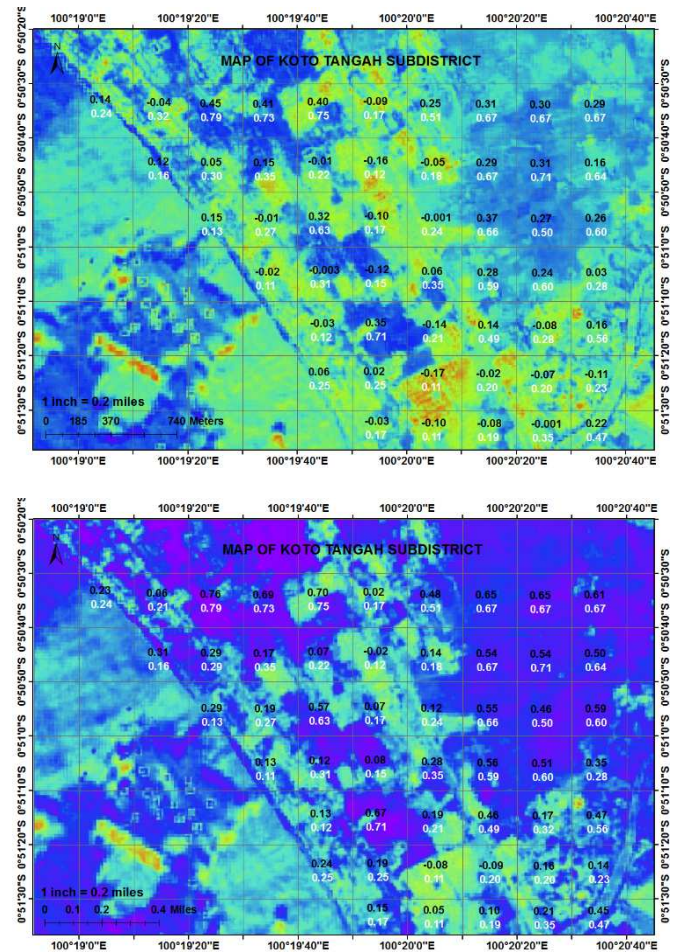

Fig. 3  The Map of water content and vegetation index in Koto Tanggah (2) subdistrict: (a). NIR-SIWR1 (b). NIR-SWIR2

Fig. 4 shows the water content and vegetation index map for the Koto Tanggah sub-district. The area has been extensively developed with residential settlements and several offices. Perupuk village has a high-water content and low vegetation along the coastline. However, the water content value decreases further away from the beach. Another location with high water content is in the Tabing airport area.

Fig. 5 shows the Padang Utara district's water content and vegetation index. This area is mainly filled with residential housing and business areas. Generally, the areas along the coastline in this region have low water content except in open areas. However, the Nanggalo subdistrict area has high water content because open spaces and rice fields still dominate it.

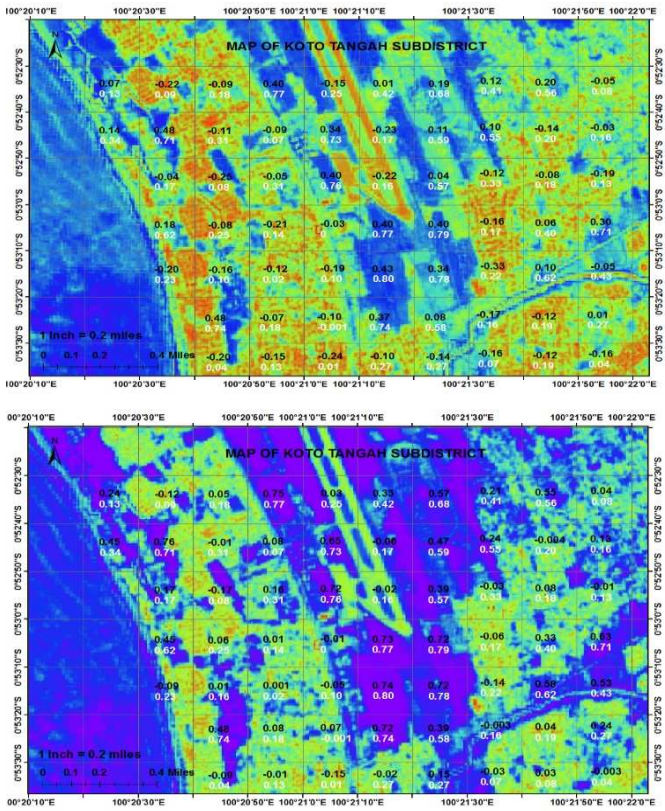

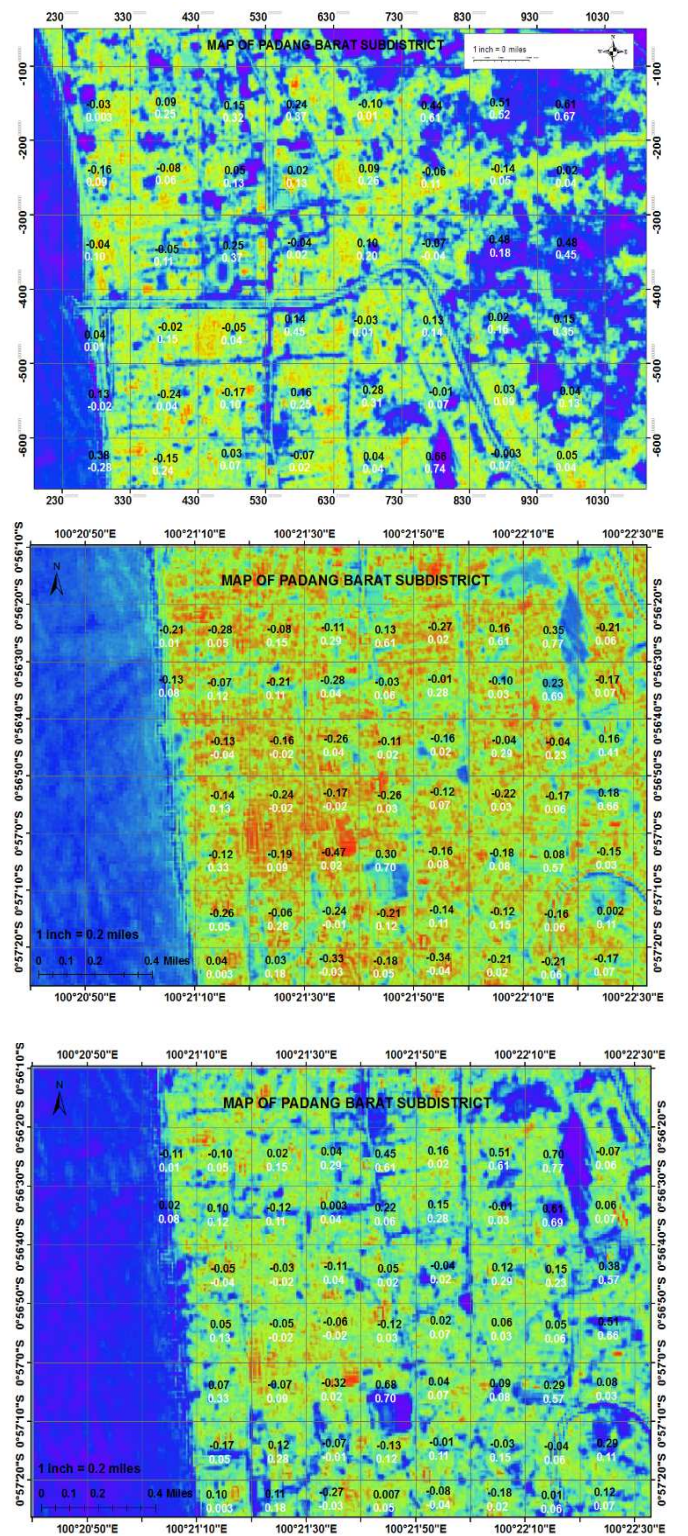

Fig. 6 depicts the western plains area's water content and vegetation index. This area is relatively densely populated, and almost every area is inhabited by residential communities. However, regions near water sources and river flows have higher water content than others. The NIR-SWIR2 wavelength, which is longer than SWIR1, allows for a clear distinction between areas with high and low water content.

Fig. 4 The Map of water content and vegetation index in Koto Tanggah (3) subdistrict: (a). NIR-SIWR1 (b). NIR-SWIR2

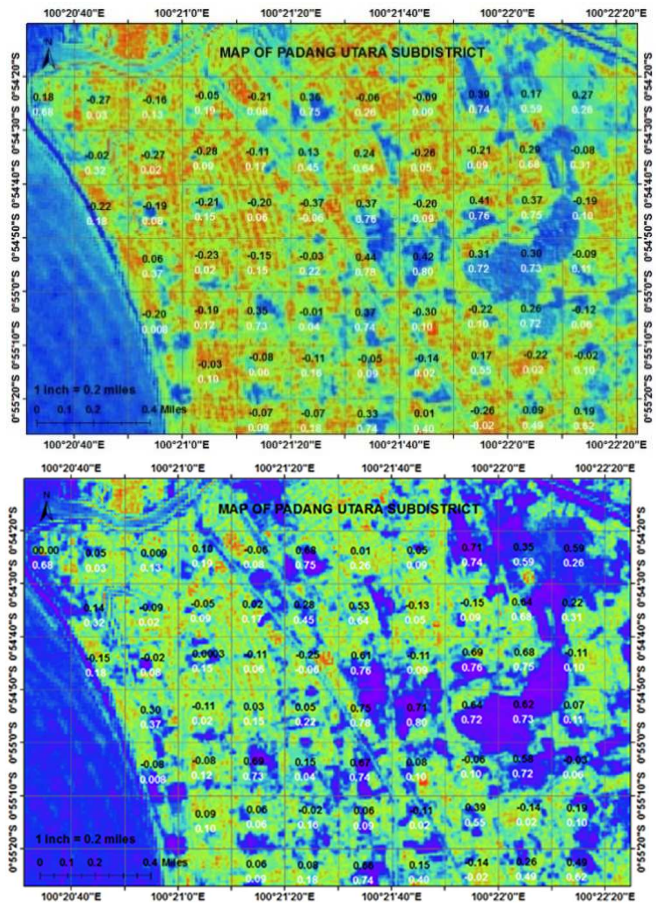

Fig. 5 The Map of water content and vegetation index at North Padang subdistrict and Nanggalo (1) subdistrict: (a). NIR-SIWR1 (b). NIR-SWIR2

Fig. 6 The Map of water content and vegetation index at West Padang subdistrict and Nanggalo (1) subdistrict: (a). NIR-SIWR1 (b). NIR-SWIR2

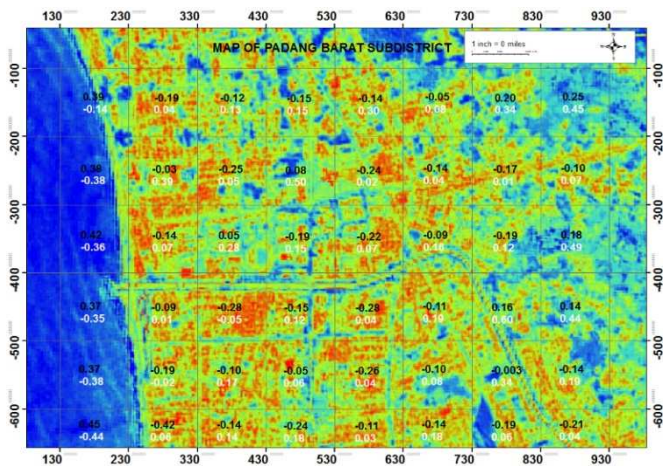

Fig. 7 The Map of water content and vegetation index at west and East Padang subdistrict (1) subdistrict: (a). NIR-SIWR1 (b). NIR-SWIR2.

Fig. 7 shows the water content and vegetation index for the West Padang and East Padang sub-districts. This area is also very dense. However, some areas around the river and low-lying regions appear to have higher water content than others. On the other hand, the eastern plains have high water content due to many open spaces and swamps.

## IV. Conclusion

This study demonstrates the efficacy of water index mapping, with the Normalized Difference Water Index (NDWI) serving as a critical metric for determining liquefaction susceptibility in Padang City. NDWI used optical satellite imaging to successfully identify water-saturated areas, including water bodies and high-moisture zones, which are essential determinants in liquefaction susceptibility. The combination of NDWI results with geological and seismic data enabled a full assessment of liquefaction-prone locations, yielding valuable insights into the geographical distribution of high-risk zones.

The findings show a substantial relationship between water saturation and liquefaction vulnerability, underlining the importance of water content as a driving force in soil instability during seismic events. This method provides a cost-effective and efficient solution to large-scale analysis, eliminating the need for costly field surveys while retaining reliability. Furthermore, the study lays the groundwork for better-informed urban planning, risk reduction techniques, and disaster preparedness initiatives in Padang City.

The NDWI-based approach is replicable and adaptable to similar geological environments, providing a scalable strategy for identifying liquefaction-prone zones in other countries. Future studies could benefit from using additional variables, such as soil type and groundwater depth, to improve vulnerability evaluations [20]. Overall, this work emphasizes the value of remote sensing and geospatial analysis in addressing geological hazards [19], improving resilience and risk reduction in earthquake-prone locations such as Padang City.

## Acknowledgment

We thank Universitas Negeri Padang for funding through the International collaborative research skim of PNBP contract number 1369/UN.35.15/LT/2023, GEE, the Japan Aerospace Exploration Agency, the United States Geological Survey (USGS), and the European Space Agency (ESA) for supporting satellite data.